\documentclass{svmult}
\usepackage{makeidx} 
\usepackage{graphicx}
\usepackage{multicol}
\usepackage{listings}
\usepackage{url}
\makeatletter
\def\url@leostyle{%
  \@ifundefined{selectfont}{\def\UrlFont{\sf}}{\def\UrlFont{\small\ttfamily}}}
\makeatother
\usepackage{amsmath}
\usepackage{amssymb}
\usepackage{bm}

\makeindex       

\begin{document}

\title*{\large Political protest Italian-style: The dissonance between the blogosphere and mainstream media in the promotion and coverage of Beppe Grillo's V-day\thanks{Postprint. \textbf{Cite as:} Alberto Pepe and Corinna di Gennaro. Political protest Italian-style: The blogosphere and mainstream media in the promotion and coverage of Beppe Grillo's V-day. \textit{First Monday.} Volume 14, Number 12. 2009}.}
\titlerunning{Political protest Italian-style}
\author{Alberto Pepe \and Corinna di Gennaro}

\institute{Alberto Pepe \at Center for Embedded Networked Sensing and Department of Information Studies, University of California, Los Angeles
\and Corinna di Gennaro \at Berkman Center for Internet \& Society, Harvard University}

\maketitle

\abstract{\footnotesize We analyze the organization, promotion and public perception of V-day, a political rally that took place on September 8, 2007, to protest against corruption in the Italian Parliament. Launched by blogger Beppe Grillo, and promoted via a word of mouth mobilization on the Italian blogosphere, V-day brought close to one million Italians in the streets on a single day, but was mostly ignored by mainstream media. This article is divided into two parts. In the first part, we analyze the volume and content of online articles published by both bloggers and mainstream news sources from June 14 (the day V-day was announced) until September 15, 2007 (one week after it took place) . We find that the success of V-day can be attributed to the coverage of bloggers and small-scale local news outlets only, suggesting a strong grassroots component in the organization of the rally. We also find a dissonant thematic relationship between content published by blogs and mainstream media: while the majority of blogs analyzed promote V-day, major mainstream media sources critique the methods of information production and dissemination employed by Grillo. Based on this finding, in the second part of the study, we explore the role of Grillo in the organization of the rally from a network analysis perspective. We study the interlinking structure of the V-day blogosphere network, to determine its structure, its levels of heterogeneity, and resilience. Our analysis contradicts the hypothesis that Grillo served as a top-down, broadcast-like source of information. Rather, we find that information about V-day was transferred across heterogeneous nodes in a moderately robust and resilient core network of blogs. We speculate that the organization of V-day represents the very first case, in Italian history, of a political demonstration developed and promoted primarily via the use of social media on the web.}

\normalsize
\section{Introduction}
\setcounter{footnote}{0}

Very early in the morning of what would be a sunny Saturday in late summer 2007, the first red ``V" appeared in Bologna's Piazza Maggiore. By early afternoon, thousands of matching ``Vs" appeared across t-shirts, flags, and banners, crowds filling the piazza and the surrounding streets of Bologna's historical downtown (Figure 1). With close to one hundred thousand people gathered in Piazza Maggiore, Bologna was V-day's epicenter, but by no means alone: parallel protests took place simultaneously in the piazzas of over 200 cities across the peninsula, involving an estimated total of about one million people\footnote{The exact number of participants in the demonstrations of September 8 was cause of much controversy. The day of the event, different sources estimated the number of participants in Bologna to be as high as 200,000 and as low as 30,000. In the days that followed, the majority of analyzed sources cite figures ranging between 50,000 and 100,000 for Bologna, and between 500,000 and 1.5 million for the event as a whole (in Italy and abroad).}. All gathered to protest corruption in Italian politics and to collect signatures demanding a new law that would ban convicted criminals from Parliament. Though drawing inspiration from the insurgent themes of the comic book and film \emph{``V for Vendetta"}, the V-day of 8 September 2007 received a decidedly Italian twist: its titular ``V" standing not for ``Vendetta" or ``Victory", but rather for ``Vaffanculo" --- Italian for ``fuck off". Vulgar, yet passionate and colorful, the name spoke to the overall tone of the demonstration, as well as to the vulgar politics being protested and the protesters' anger over a political system that was blatantly and fatally flawed.

\begin{figure}[h!]
	\centering
		\includegraphics[width=1.0\textwidth]{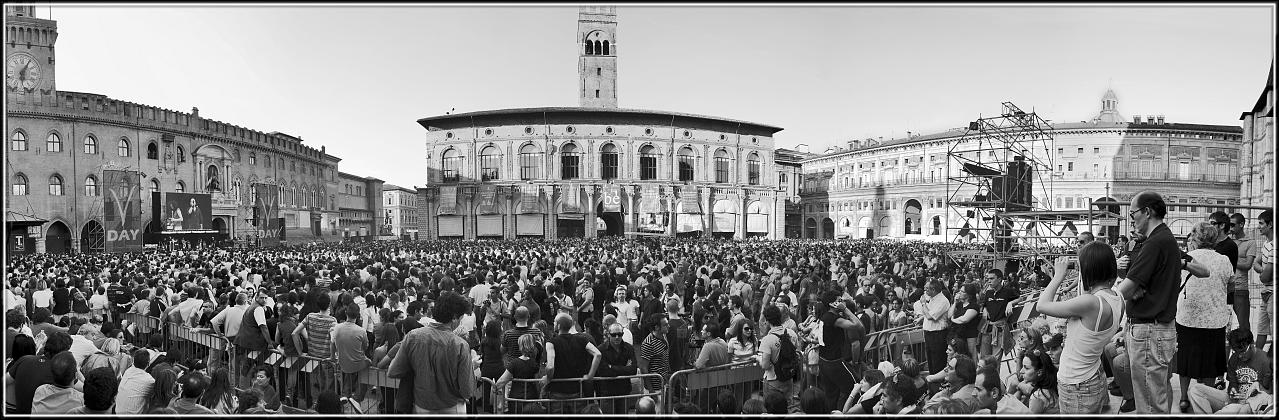}
	\caption{Piazza Maggiore (Bologna) during V-day rally, September 8, 2007 (Photo by Flickr user paPisc)}
	\label{fig:1}
\end{figure}

To analyze both the motives behind such a large-scale protest and its method of organization at the grassroots level, it is important to look at contemporary forms of political organization and participation, in general, and, more specifically, on how they are developing in the contemporary Italian media and political landscape. An illustrative example of the interplay between the protests and the media can be found by asking: how did major daily newspapers cover V-day? Figure 2 displays the front pages of the two most circulated Italian national daily newspapers, \emph{Il Corriere della Sera} (also known as \emph{Il Corriere}) and \emph{La Repubblica} on 8 September 2007\footnote{\emph{Il Corriere della Sera} and \emph{La Repubblica} are the most circulated daily newspapers in Italy with average circulation of 681,854 and 630,080, respectively, as of 28 May 2009. Source: Wikipedia at \url{http://it.wikipedia.org/wiki/Quotidiani}, accessed 28 May 2009.}. 

\begin{figure}[h!]
	\centering
		\includegraphics[width=1.0\textwidth]{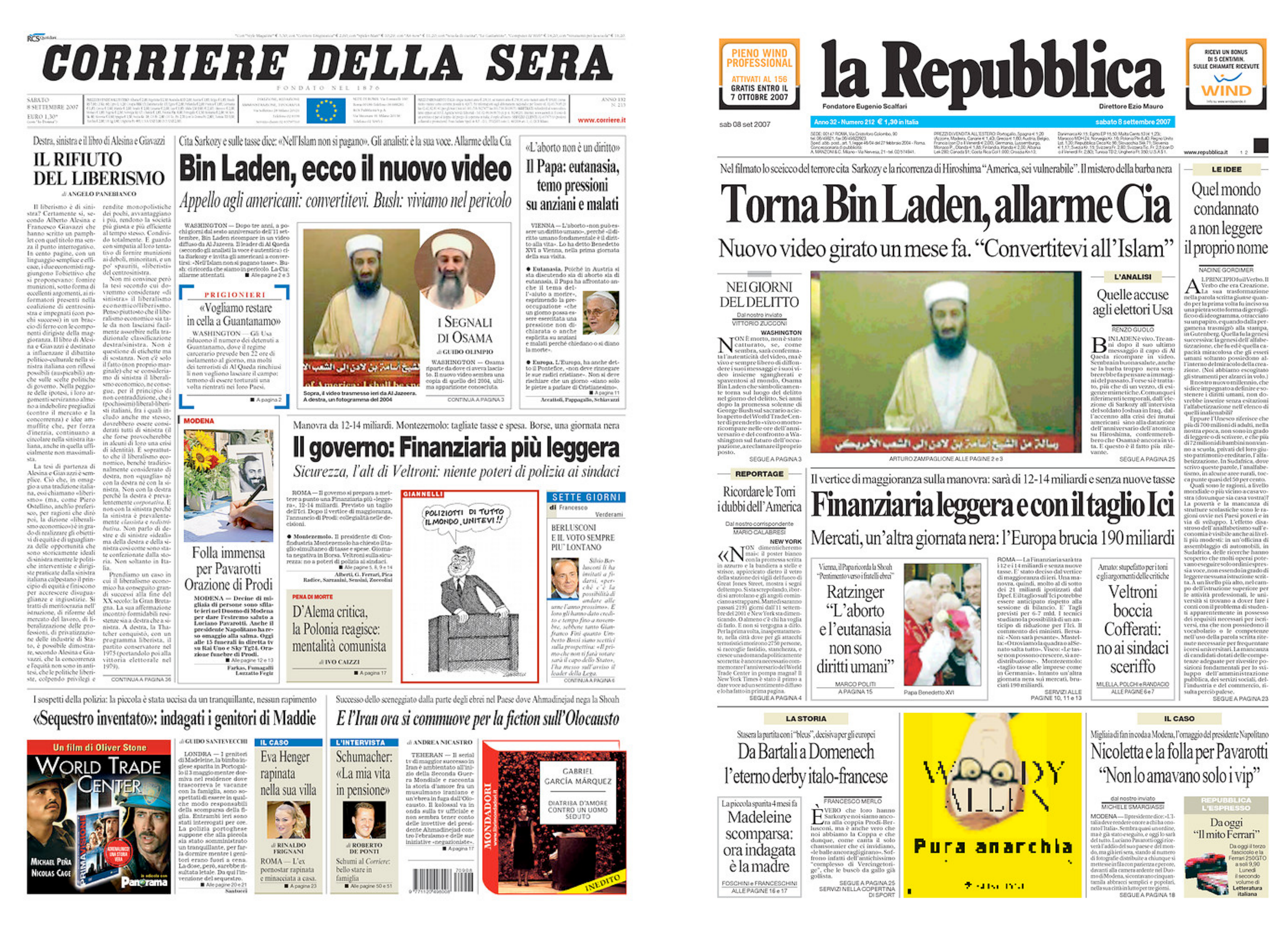}
	\caption{ Front pages of Italian daily newspapers Corriere della Sera and La Repubblica on September 8, 2007}
	\label{fig:2}
\end{figure}

Their front pages are not very dissimilar. Leading stories include a potential return onto the international stage of Osama Bin Laden, the funeral ceremonies honoring tenor Luciano Pavarotti (who had died two days before), and the position of the Vatican on abortion and euthanasia. Also finding a home on \emph{Il Corriere's} homepage are stories of a recent burglary at the villa of a former adult film actress and on the retired life of a former Formula One driver, while \emph{La Repubblica} reports on the upcoming soccer match between Italy and France. Nowhere on these pages is there a single mention of V-day, a political rally that brought in a single day one million Italians to the streets. A closer examination of the inside pages of these papers reveals that V-day was not simply under-reported by the mainstream media: it was ignored almost entirely. This preliminary finding motivated the study presented in this article.

V-day was launched on the personal blog of popular Italian comedian and activist Beppe Grillo on 14 June 2007, igniting a word-of-mouth announcement of the event across the Italian blogosphere. During the following months, V-day was promoted, organized, and discussed both online and offline by fellow bloggers, political activists, volunteers in local MeetUp.com groups, a handful of journalists and politicians, and many regular citizens. On 8 September, people took to the streets to protest against corruption in Italian politics. In a single day, over 330,000 physical signatures were collected to demand a referendum for a new law banning convicted criminals from Parliament (in Italy, the number of signatures required for a referendum is 50,000). For Italians, it was the first experiment in political organization and action developed primarily via the use of social media on the Web.

The online organization and the subsequent off-line mobilization of such a large-scale public demonstration took place largely without the promotion, influence, support of or criticism by the mass media, which is an unprecedented event in contemporary Italian history. As we will demonstrate throughout this article, in spite of its popularity in the streets, the event was largely ignored by the mainstream media, before, during and after it took place. This fact alone makes V-day an interesting case study to understand the relationship between mainstream media, represented by traditional news sources, and social media, in the form of blogs and similar user-generated content platforms on the Internet, in bringing about political action.

In this article, we investigate the role played by social and mass media in promoting and organizing V-day by asking these research questions: 1) how was this event organized? 2) what role did Web-based social media play in the organization of this political demonstration? 3) how was the event reported, promoted or criticized by online news sources? 4) how robust and connected is the network of bloggers responsible for the organization of the event? To address these questions, we employ the entire record of personal blogs and online news sources about V-day published in Italian in the period from 14 June (the day V-day was announced) until 15 September 2007 (a week after it took place). We perform two types of analysis. First, we explore the ways by which the blogosphere and mainstream media covered the event. We measure the volume of items published by both bloggers and news sources about V-day and determine, via a content analysis, whether they favor, oppose or are neutral to the demonstration. Second, we explore the methods of information production and dissemination that led to the rally, by studying the interlinking structure of the V-day blogosphere network and its levels of heterogeneity and resilience.

This article is organized as follows. In the next section, we provide an overview of the Italian political and media landscape by linking it to existing research in this area. In the following sections, we present methods and findings of our study in two parts. Part 1 presents the analyses of blogs and news sources. Part 2 presents the results from the network analysis of the V-day blogosphere. We conclude with a brief discussion of our results.

\section{Beppe Grillo's blog, Italian media ecology and political activism}
Beppe Grillo is a comedian-turned-political activist. In the early 1990s, Grillo was banned from Italian public broadcasting television after making fierce satirical comments against the ruling Italian Socialist Party. In October 2005, Grillo started a personal blog, Beppegrillo.it, and started blogging daily about current affairs and political events, with a strong emphasis on freedom of information, environmental sustainability, political corruption, and citizen activism. In the past three years, the blog has reached a remarkable level of readership and active discussion. Beppegrillo.it is not only Italy's most read and discussed blog, it was recently ranked by the \emph{Guardian} among the top 10 most influential and most visited blogs in the world, alongside popular English-language political blogs such as the Huffington Post and Talking Points Memo\footnote{The standing and influence of blogs in the blogosphere are calculated daily and published by Technorati (\url{http://www.technorati.com/}). The world's 50 most powerful blogs, based on Technorati rankings, were published on the \emph{Guardian} in March 2008. List available at \url{http://www.guardian.co.uk/technology/2008/mar/09/blogs}, accessed 10 June 2009}.

The phenomenon of Beppe Grillo's blog popularity exemplifies the power of the Internet as a source of uncensored information and as a place for the free discussion and sharing of ideas regardless of political partisanship. As such, it is of particular importance in a country like Italy, which has a peculiar media ecology when compared to other Western democracies, with a mainstream media monopoly held firmly in the hands of current Prime Minister Silvio Berlusconi. The three Italian public broadcasting channels from Radiotelevisione Italiana (RAI 1, RAI 2, and RAI 3) are state-owned and its board members are elected by political parties (see Navarria, 2009 for a detailed overview). Berlusconi's family also owns Mediaset, Italy's largest commercial television group, with three television channels (Italia 1, Retequattro, Canale 5), in addition to a daily newspaper magazine (\emph{Il Giornale}) and the country's biggest publishing house (Mondadori Group) that controls a large portion of Italian and international news magazines (including \emph{Panorama, Grazia, Focus, Cosmopolitan}, and \emph{TV Sorrisi \& Canzoni}). In such a media landscape, the Web has become an important alternative medium to the highly politicized and biased traditional media sources. The Italian blogosphere is awash with media content that reports current affairs and political actions, which are left untold or entirely ignored by the mainstream media. These stories are not only being reported by regular citizens, but also by established journalists and intellectuals who have made the Internet one of their preferred channels of communication to the wider public. Besides Grillo, examples are journalist Marco Travaglio, comedian Daniele Luttazzi, writer Jacopo Fo, and theatre actress Franca Rame, all of whom appear in the results of this study.

The importance of social and political participation through Web-based media is most evident in the case of Grillo's blog. Not only has the blog itself become an open forum for political discussion, it has elicited novel forms of collective action via dedicated MeetUp.com groups, political demonstrations and other similar forms of protest (Navarria, 2008; 2009). In other words, Grillo's blog has acted as a lever to transfer online forms of participation and discussion to \emph{off-line} realms: \emph{from the blogosphere to the streets}. Beppe Grillo's blog was particularly successful in achieving this transference on an unprecedented scale, both on a national and international level. Since long before the September 2007 V-day demonstrations, Beppe Grillo's blog has been acting as an online facilitator for the gathering of off-line social networks of like-minded people, organized via the online portal MeetUp.com\footnote{As of 1 June 2009, The Beppe Grillo MeetUp, at \url{http://beppegrillo.meetup.com/}, counts 435 groups worldwide, 77,064 members, in 322 cities spanning across 17 countries}. These dynamics of online organization shifting into political collective action are evident in the recent V-day protest. On 14 June 2007, V-day --- a public demonstration aimed at removing from office members of the Italian Parliament with criminal convictions --- was announced on Beppe Grillo's blog. The announcement was followed by an extraordinary level of public engagement and mobilization which took place both \emph{online}, in the form of comments on Beppe Grillo's blog, blog trackbacks, online videos and music, and \emph{off-line}, via the organization of talks, open forums, and meet-ups. While we do not investigate the mobilization processes that took place off-line amongst MeetUp.com volunteers (who were essential in organizing and coordinating the events locally in the different Italian towns), the focus of our study lies in the online mobilization and recruitment strategies via blogs of journalists and non-professionals alike.

The power of the Internet to act as a means of political mobilization has been well documented in the literature. Studies of election campaigns both in Europe and America have shown how both political parties and politicians have used the Internet for their campaign efforts (Best and Krueger, 2005; Gibson, \emph{et al.}, 2005; Johnson and Kaye, 1998; Tolbert and McNeal, 2003). In addition, scholars of the Internet such as Rheingold (2002) and Shirky (2008) have looked at the role played by information and communication technologies in bringing about collective action, through the aiding and promoting of the action of regular citizens. Rheingold uses the term `smart mobs', while Shirky talks about the `power of organizing without organizations'. In addition, several case studies have looked at the effect of the Internet and mobile technologies in facilitating the gathering together of people for political protest, for example during the 2004 Ukraine Orange revolution (Goldstein, 2007) and the 2007 Burma Saffron revolution (Chowdhury, 2008). Studies have also looked a the role played by citizen journalism in influencing elections, for example the case of the OhMyNews site in South Korea in 2002 elections (Joyce, 2007). In addition, many of the case studies cited above show that the effect of the Internet on democracy cannot be analyzed by itself, but instead must (i) take into consideration the interplay between online and off-line practices and (ii) be analyzed in the context of a country's media ecology. Scholars of social movements have already applied this theoretical framework in a fruitful way. For instance, Andretta, \emph{et al.} (2002) demonstrated, through a quantitative survey and interviews with activists, that rather than competition between online and off-line activities, they rather frequently match and work in parallel within social movements.

A great deal of recent literature is concerned specifically with online content production, and the analysis of the dynamic relationship between emerging citizen and professional journalism. Reese, \emph{et al.} (2007), for example, analyze top U.S. political blogs and their thematic relationship with the traditional professional media, searching for the presence of an echo-chamber effect, \emph{i.e.}, the idea that the content of political blogs simply ``echoes" the message and ideology of selected mainstream media. Kevin Wallsten (2005) performs a similar study on a larger population set, exploring the media echo-chamber effect on blogs with low readership, authored by average citizens. In other work, the relationship between bloggers and professional journalism has been noted to be much less harmonious, with blogs challenging mainstream media, as in the case of the photoshopped images from Lebanon published by Reuters during the Israel-Hezbollah War of 2006 (Usher, 2008).

While the use of technology has been successful in all these case studies in either facilitating or allowing crowd mobilization, the collective action aided by the technology has not always succeeded in bringing about political change. While some may argue that the effect of the Internet on collective action is only successful if political change is actually achieved, what we focus on in this article is strictly the role that the Italian blogosphere played in the promotion and organization of V-day, regardless of their subsequent impact on the political system. What is remarkable about the organization of V-day is that it received virtually no coverage by the mainstream media. Yet, its three-month preparation, as well as its day-long activities that took place across the nation, were largely documented on dedicated online video channels, photo repositories, blogs, and other Web-based user-generated media. This is the story that this article sets out to investigate.

\section{Part 1: Content analysis --- The blogosphere vs. mainstream media}

\subsection{Data and methods}

Both blog and online news archives are becoming important information sources for a growing body of Internet-based research. Despite the increasing popularity and academic relevance of these sources, dedicated search engines, portals and aggregators of blog and news data are still relatively scarce and, very importantly, they lack tools and interfaces for large-scale analyses. We employed a recent comparative review of blog search engines (Thelwall and Hasler, 2007) to assess the coverage and reliability of blog search engines for our study. Although we could not find a similar comparative review for news archive platforms, we performed some light evaluation tests of notable search engines for news, such as Google News Archive Search and Yahoo News Search, as well as an academic project, Columbia University's Newsblaster (Barzilay, \emph{et al.}, 2002).

For the purpose of the present study, our main search platform requirements can be summarized as follows: (i) the possibility to search and classify the content of individual news items and blog posts (\emph{i.e.}, not just the overall content of news sites and blogs); (ii) the possibility to search at a day-level granularity for the time period under study (\emph{e.g.}, a typical query could be: ``search all the blog posts written on 8 September 2007"); (iii) the possibility to filter through news items and blog posts written in a specific language, \emph{i.e.}, Italian;\footnote{We decide to ignore, in this article, articles published in the foreign press, despite the fact that leading stories appeared on popular newspapers both before and after V-day. The site \url{http://vday.wordpress.com/} contains links to articles appeared in the month of August and September on the \emph{Independent, International Herald Tribune, La Vanguardia, El Pais, Le Figaro}, and \emph{Le Monde}} and, (iv) the possibility of harvesting all relevant results in some structured (or semi-structured) format. With these requirements in mind, after some evaluation tests we decided to use the Google News Archive Search\footnote{Google News Archive Search, at \url{http://news.google.com/archivesearch}, accessed 10 July 2009} and the Google Blog Search\footnote{Google Blog Search, at \url{http://blogsearch.google.com/}, accessed 10 July 2009.}, to find and collect news items and blog posts, respectively.

Given the vast amount of data to be analyzed (three months worth of blog and news data), harvesting the full archives for the entire period and performing an analysis of relevant blog posts and news items \emph{a posteriori} was not feasible. Instead, we decided to rely upon the search algorithm of these platforms to pre-select blog posts and news items about V-day. For this reason, the fact that both archives employed use the same underlying search algorithm (provided by Google Search) was deemed an advantage as it strengthened the validity and reliability of our findings.

The news and blog archives were searched using the following query terms:

\lstset{basicstyle=\scriptsize}
\begin{lstlisting}[frame=lines]
v-day OR v day OR vday OR vaffaday OR vaffanculo day
\end{lstlisting}

for every day in the three-month period from 14 June until 15 September 2007. The aim of such a broad search query was to attempt to gather a corpus of all possible blog items and news posts related to V-day, and subsequently subject the corpus to manual filtering. Every result obtained was manually checked, \emph{i.e.}, every URL of each blog post and news item in the collected corpus was resolved and its page content was analyzed. This manual analysis was performed to both (i) assess the validity of every result obtained and (ii) categorize the content of the blog post or news item as \emph{in favor, opposed} or \emph{neutral} to V-day. Table 1 presents a summary of the results obtained through this manual filtering process.

\begin{table*}[htp]
\centering
\begin{footnotesize}
\begin{tabular*}{\textwidth}{@{\extracolsep{\fill}}lrr}
Results&\textbf{Blog posts}&\textbf{News items}\\
\hline\hline
Valid&802 (43\%)&196 (44\%)\\
Discarded&1038 (57\%)&246 (56\%)\\
\hline
Total returned&1840&442\\
\end{tabular*}
\end{footnotesize}
\caption{Table 1: Breakdown of the search results obtained (blog posts and news items related to V-day)}
\end{table*}

Table 1 displays the total number of blog posts and news items returned by Google Blog Search and Google News Archive Search (1,840 and 442, respectively) and the number of results that were deemed valid after manual checking (802 and 196, respectively). This reveals that over half of the results obtained were discarded from the blog and news corpora, for a number of different reasons.

A result in the blog corpus was considered valid if the blog post (i) is primarily about V-day or mentions V-day in a significant way, (ii) promotes, criticizes, or reports V-day, (iii) contains an active discussion about V-day that engages the blog post author, as well as other blog readers. A number of results returned by Google Blog Search had to be discarded from the corpus for a variety of reasons including: (i) V-day was found only as a tag, a banner or a link in the blog sidebar, header or footer; (ii) V-day was discussed in the comment section of the blog post, but without engaging the blog post author\footnote{The blogs of Antonio Di Pietro and Piero Ricca are typical examples of blogs which received a constant and substantial stream of reader comments about V-day in the period under study. Although Di Pietro and Ricca eventually blogged about V-day, they did not actively engage in discussion with the readers in many occasions. For this reason, a number of blog posts by Di Pietro and Ricca were discarded. Yet, it is interesting that their blogs acted as platform for active discussion about V-day among their readers.}; and, (iii) blog post was not available anymore. Also, blog posts authored by Beppe Grillo, on beppegrillo.it, were discarded. Finally, in some cases we found blog posts about V-day that were authored by professional journalists on blog-like platforms hosted by a major newspaper. In these cases, results were removed from the blog corpus and moved to the news corpus.

A result in the news corpus was considered valid if the news item mentioned V-day. Unlike blogs, most news sites do not enable comments and discussion by readers and thus, none of the results in the news corpus was found to have significant reader conversations concerning V-day. The majority of results discarded from the news corpus were duplicates, \emph{i.e.}, identical news items hosted at multiple sites and/or at news aggregator Web sites. Also, in some cases we found that some of the results in the news corpus were in fact blog posts, \emph{i.e.}, they were hosted on a blog platform, authored by non-professional journalists and written in a very informal, blog-like style. In these cases, these results were moved to the blog corpus. These borderline cases (blog posts that appear in news results and news items that appear in blog results) are not just technical inconsistencies. They reveal that the boundary separating blogs from news is thin and blurry. This point is discussed in greater detail later in the article.

Besides assessing the validity of the entries in the news and blog corpora, we also performed a light content analysis to determine whether news items and blog posts were in favor, opposed or neutral to the V-day rally. It is important to note that, in order to perform this classification, we considered every entry independently of the overall political inclination of blogs and media news sites. The classification was performed based on the following guidelines. Results were classified as ``in favor" if the author of the blog post or the news item explicitly endorses the political rally or if the support is clear from the overall tone and argument presented. Results were classified as ``opposed" if the author of the blog post or news item is heavily critical of the political rally. Many bloggers opposed to the rally, for example, explicitly announced that they did not endorse and would not attend the event. Results were classified as neutral when a prevailing opinion (in favor or opposed) does not emerge from reading the blog post or news item, or when the author explicitly mentions that they have no opinion or are undecided about it. We used cumulative classification results to automatically determine blogs' and news outlets' overall inclination towards V-day, \emph{e.g.}, if the majority of posts analyzed for a certain blog were found to be in favor then the entire blog was classified as in favor. Borderline cases, \emph{e.g.}, blogs containing items with contrasting views towards V-day and blogs for which not enough material was available, were assessed manually.

\subsection{Findings}

Results from our content analysis of blog posts and news items about V-day are presented in Table 2. We find that the number of valid blog post results exceeds that of news items by a factor of four (802 and 196 results, respectively), indicating that V-day was covered more prominently by the blogosphere than by mainstream news media. Also, the majority of blog posts analyzed were in favor of V-day (78 percent), whereas analyzed online media sources were roughly evenly split across news items in favor of and neutral to the political rally (41 percent and 49 percent, respectively), indicating that the blogosphere was overall more supportive of V-day compared to mainstream media sources.

\begin{table*}[htp]
\centering
\begin{footnotesize}
\begin{tabular*}{\textwidth}{@{\extracolsep{\fill}}lrr}
Results&\textbf{Blog posts}&\textbf{News items}\\
\hline\hline
\textit{In favor}&628 (78\%)&81 (41\%)\\
\textit{Neutral}&96 (12\%)&96 (49\%)\\
\textit{Opposed}&78 (10\%)&19 (10\%)\\
\hline
Total valid&802&196\\
\end{tabular*}
\end{footnotesize}
\caption{Blog posts and news items by stance towards V-day}
\end{table*}

While Table 2 gives an idea of the overall volume of stories published about V-day in the period under study, we wanted to deepen our understanding of the actual configuration of blog and media ecologies in which these stories were published. For this reason, we present, in Tables 3 and 4, the most recurrent entries in blog and news corpora, \emph{i.e.}, the blogs and news sources that most discussed V-day in the period under study.

Table 3 contains the most recurrent sites in the blog corpus, organized by their attitude towards V-day (in favor, neutral, opposed), along with the number of posts they published about V-day, the number of reactions they received (\emph{i.e.}, the number of times they were linked to by other blogs) and their Technorati ranking, as of 14 May 2009. Nearly all sites included in this list are both sole-authored and fairly low ranking in the blogosphere; thus, these are likely to be blogs with limited readership and visibility. In other words, the sites that discussed V-day more prominently (whether to support it, report it or critique it) were likely to be authored by average citizens and not by ``A-list" bloggers (also known in the blogosphere as ``blogstars" or V.I.B. --- ``Very Important Bloggers"). It follows that the bulk of the political promotion, dissemination, and critique of V-day (in volume of posts) relied on the bottom-up effort of regular citizens.

\begin{table*}[htp]
\centering
\begin{footnotesize}
\begin{tabular*}{\textwidth}{@{\extracolsep{\fill}}lccr}
\textbf{URL}&\textbf{V-day posts}&\textbf{reactions}&\textbf{rank}\\
\hline\hline
\textit{In favor}&&&\\
\hline
http://alessios4.blogspot.com/&12&1,255&17,725\\
http://www.onemoreblog.it/&10&348&-\\
http://mizcesena.blogspot.com/&8&120&248,101\\
http://laltrafacciadellamedaglia.blogspot.com/&7&22&497,196\\
http://faenzattiva.blogspot.com/&7&20&722,455\\
http://italia-tragica.blogspot.com/&6&2&-\\
http://www.francescaferrara.net/&5&52&271,914\\
http://satiradanzante.blogspot.com/&5&14&539,256\\
http://odiostudioaperto.blogspot.com/&5&449&53,759\vspace*{0.2in}\\
\hline
\textit{Neutral}&&&\\
\hline
http://www.canisciolti.info/&3&517&-\\
http://graziepertuttoilpesce.blogspot.com/&2&4&-\\
http://aghost.wordpress.com/&2&130&168,330\vspace*{0.2in}\\
\hline
\textit{Opposed}&&&\\
\hline
http://www.francescocosta.net/&4&533&31,787\\
http://www.terzoocchio.org/&2&97&316,328\\
http://www.lucaconti.it/&2&117&131,771\\
http://lampidipensiero.wordpress.com/&2&213&316,224\\
http://gianmariomariniello.it/&2&57&588,858\\
\end{tabular*}
\end{footnotesize}
\caption{Most recurrent sites in the blog post corpus by stance towards V-day}
\end{table*}

Table 4 contains the most recurrent sources in the news corpus, based on their stance towards V-day (in favor, neutral, opposed), along with their scope (nationwide or local circulation) and their format (``Y" if a print version is available, ``N" if they are only available online). All the news sources in favor of V-day are local online newspapers with scope and intended circulation that rarely goes beyond the level of Italian provinces and regions. Unlike many larger, professional mainstream houses, most of these outlets explicitly endorsed V-day. It is interesting to note that these small, local news outlets are often hybrids, for they function at the intersection of professional journalism and social media. For example, the venue with the largest number of stories about V-day (TeleFree) identifies itself as an ``institutional blog", even though it features an editorial board and regular professional contributors. Despite this fact, its readership is probably fairly limited --- similar, if not lower, than that of some political and news blogs. It is at this level that the line between blogs and traditional news sources is at its thinnest. In the ``neutral" section of Table 4, all well-established Italian news agencies can be found. The printed copies of these newspapers are widely circulated nationwide, but their Web sites are also highly popular and are among the most visited sites in Italy\footnote{As of 4 July 2009, \emph{La Repubblica} and \emph{Il Corriere della Sera} are the 11th and 14th most visited Web sites in Italy, respectively. Source Alexa, at \url{http://www.alexa.com/topsites/countries/IT}, accessed 10 July 2009}.

Finally, none of the newspapers analyzed published a considerable amount of stories against V-day, but among the few entries in the lower portion of Table 4, one finds \emph{Il Giornale}, Tg Mediaset and \emph{Sorrisi e Canzoni} --- two news sources and a tabloid, respectively, owned by the Berlusconi family. In sum, only small, local newspapers explicitly and openly endorsed V-day. Most larger mainstream news outlets, whatever their political orientation, were very careful to maintain a neutral stance towards V-day, aside from the Berlusconi-owned media outlets which were openly opposed. This scenario of media control is not particularly surprising given the current state of affairs with Italian media and the media's collusion with politics.

\begin{table*}[htp]
\centering
\begin{footnotesize}
\begin{tabular*}{\textwidth}{@{\extracolsep{\fill}}lccr}
\textbf{URL}&\textbf{V-day items}&\textbf{scope}&\textbf{print}\\
\hline\hline
\textit{In favor}&&&\\
\hline
http://www.telefree.it/&6&Local&N\\
http://www.altromolise.it/&4&Local&N\\
http://www3.varesenews.it/&3&Local&N\\
http://www.tranionline.it/&3&Local&N\\
http://www.pupia.tv/&3&Local&N\\
http://www.isolapossibile.it/&3&Local&N\\
http://www.intoscana.it/&3&Local&N\\
http://www.brundisium.net/&3&Local&N\\
http://www.viveresenigallia.it/&2&Local&N\\
http://www.traniweb.it/&2&Local&N\vspace*{0.2in}\\
\hline
\textit{Neutral}&&&\\
\hline
http://www.repubblica.it/&9&Nationwide&Y\\
http://www.lastampa.it/&8&Nationwide&Y\\
http://www.corriere.it/&8&Nationwide&Y\\
http://quotidianonet.ilsole24ore.com/&6&Nationwide&Y\\
http://ilrestodelcarlino.ilsole24ore.com/&3&Nationwide&Y\\
http://espresso.repubblica.it/&3&Nationwide&Y\\
http://bologna.repubblica.it/&3&Local&Y\\
http://www.tendenzeonline.info/&2&Nationwide&N\\
http://www.skylife.it/&2&Nationwide&N\\
http://www.mentelocale.it/&2&Local&N\vspace*{0.2in}\\

\hline
\textit{Opposed}&&&\\
\hline
http://www.affaritaliani.it/&3&Nationwide&N\\
http://www.tgcom.mediaset.it/&2&Nationwide&N\\
http://www.ilgiornale.it/&2&Nationwide&Y\\
http://www.aprileonline.info/&2&Nationwide&N\\
http://www.sorrisi.com/&1&Nationwide&Y\\
\end{tabular*}
\end{footnotesize}
\caption{Most recurrent sites in the news item corpus by stance towards V-day}
\end{table*}

After this preliminary analysis, we decided to unfold our content analysis longitudinally, to produce a more distilled picture of the dynamic interplay, or the lack thereof, between the blogosphere and the news media over time. In this context, it is worth making a methodological remark. We performed our content analysis following the temporal structure of our corpus, \emph{i.e.}, we analyzed news items and blog posts from the announcement of V-day (14 June) to the end (15 September). By doing this, we were able to analyze the content of blog posts and news items independently, but, at the same time, we were able to place them in the context of a narrative in development, characterized by a timely, dynamic relationship between the events, the themes and the opinions discussed on the blogosphere and news media. For this reason, we present our findings below as time-series charts, and we discuss them following a narrative structure, in an attempt to appropriately reconstruct the timeline and the unfolding of the events.

The results of our longitudinal content analysis of 802 blog posts and 196 news stories about V-day for the period from 14 June to 15 September 2007 are summarized in Figure 3. The three diagrams depict the volume of blog posts and news items published over time. The top chart in Figure 3 presents the volume of blog posts (solid grey line) vs. news items (dashed blue line). The chart has been annotated with some major events that we gathered during our content analysis of the blog and news corpora. The center and bottom time-series charts of Figure 3 depict, respectively, the volume of blog posts and news items about V-day, for the period under study. The color of the area under each line indicates whether blog posts and news items were in favor, neutral or against V-day.

\begin{figure}[htp!]
	\centering
		\includegraphics[width=0.95\textwidth]{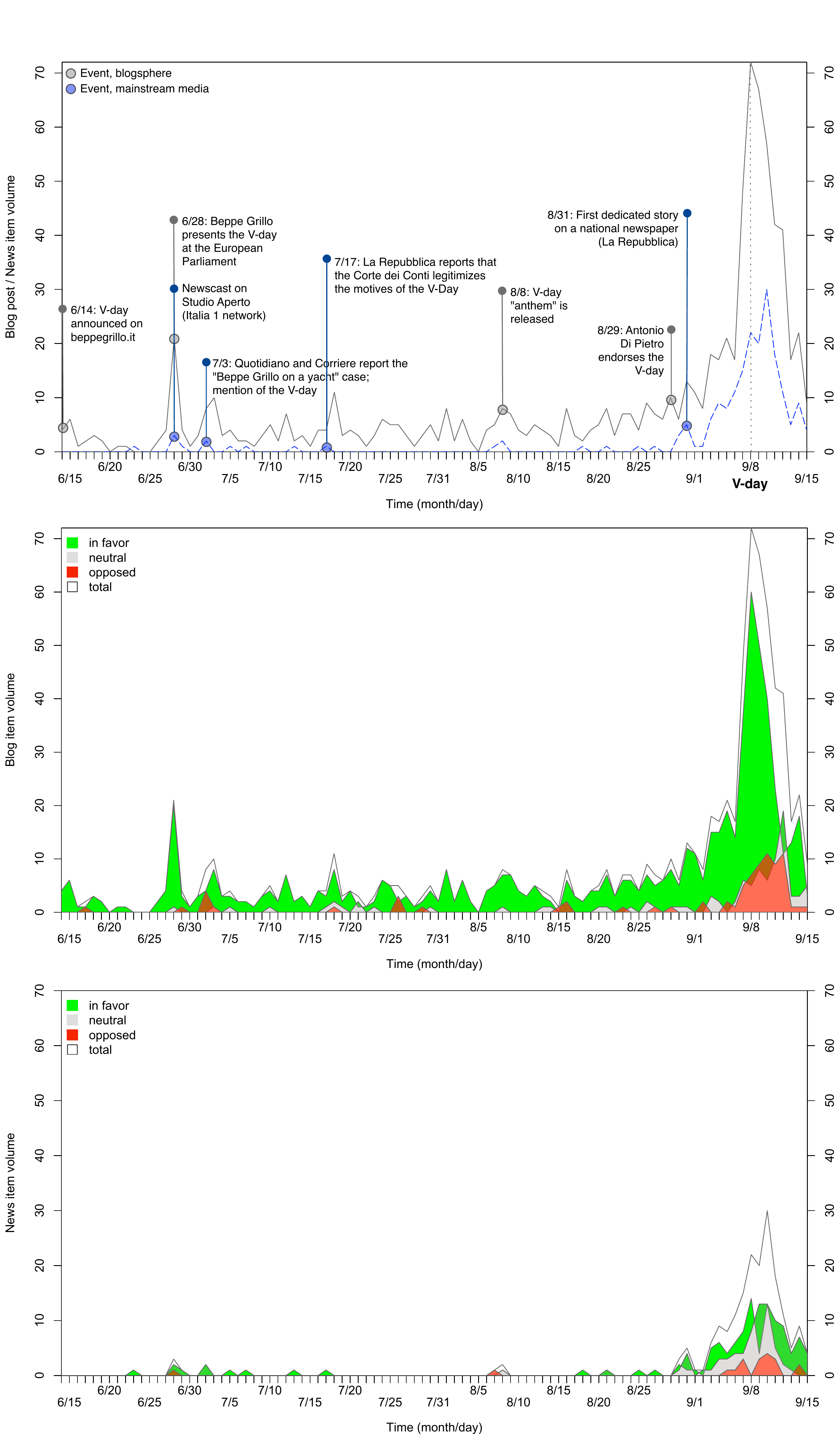}
	\caption{Time series charts depicting volume of blog posts and news items discussing V-day from 14 June to 15 September 2007: total volume chart with event annotations (top), blog post volume chart (center), news item volume chart (bottom)}
	\label{fig:3}
\end{figure}

V-day was announced with a blog post on Beppe Grillo's blog on 14 June 2007\footnote{Blog post announcing V-day, at \url{http://www.beppegrillo.it/2007/06/vaffanculoday/index.html}, accessed 10 July 2009.}. In the blog post containing the announcement, V-day was defined as an amalgam between D-Day (the Normandy Landings) and a ``spirit of Vendetta" (along with the text of the blog post comes an image of vigilante V, from the movie and comic book \emph{``V for Vendetta"}). In the following two weeks, up to 28 June, the Italian blogosphere reacted lightly to the announcement of the event --- just over 30 blogs reported or commented on it --- but overwhelmingly positively. In the mass media realm, in this period V-day was only mentioned by small-scale news outlet TeleFree (on 23 June).

On 28 June, there was a sudden increase in blogging activity about V-day, with over 20 blog posts promoting the event, published in a single day. We find that this sharp rise of blog posts for this day can be attributed to Beppe Grillo's speech at the European Parliament a day earlier. During the speech, Beppe Grillo presented the motivation behind the V-day rally and formally announced it. Beppe Grillo summarized the speech in a blog post\footnote{Blog post titled ``Parlamento Europeo e V-day", at \url{http://www.beppegrillo.it/2007/06/parlamento\_euro\_1/index.html}, accessed 10 July 2009} that also included a video that many bloggers re-published on their personal blogs. Studying in detail the content of the blogs posted on this day, we find that a small proportion of the bloggers heard about V-day for the first time on this occasion; not just from Beppe Grillo's blog, but also from a TV newscast on Studio Aperto (Italia 1) which briefly mentioned the rally, in the context of a fairly critical piece on Beppe Grillo. Also, a major national online news source (RAI News 24) briefly reported, for the first time, on the organization of the rally.

On 2 and 3 July, there was another sharp rise in the number of blog posts about V-day. Besides a moderate number (about 15) of blog posts in favor of the event, the first significant batch of blog posts critical of V-day (five, in total) appeared at this time. These blogging events corresponded with the publication of the ``polluting yacht" story by the online newspapers \emph{Il Corriere della Sera} and \emph{Quotidiano}\footnote{The ``polluting yacht" story on \emph{Il Corriere}, at \url{http://www.corriere.it/Primo\_Piano/Cronache/2007/07\_Luglio/02/grillo\_yacht.shtml}, accessed 1 July 2009. The story was previously run by tabloid magazine \emph{Eva 3000} and newspaper \emph{Libero}}. In the story, Beppe Grillo (a fervid environmentalist) was pictured aboard a yacht and criticized. V-day was briefly mentioned at the end of the article. This news story was the subject of considerable critical commentary by bloggers in favor of the political rally.

A constant stream of blogging activity in favor of V-day (with an average of five blog posts per day) characterized the period from 3 July to 16 July. A small number of news stories about V-day also appeared in this period, but a quick analysis reveals that they are all by small news agencies with local and fairly limited reach. The next rise in blogging activity (exceeding 10 blog posts in a single day) was registered on 17 July and corresponds with an article in \emph{La Repubblica} which reported that the Italian Court of Accounts (Corte dei Conti) formally legitimized the motives behind Grillo's political rally\footnote{\emph{La Repubblica} reports that the Corte dei Conti legitimizes V-day, at \url{http://www.repubblica.it/2007/07/sezioni/politica/grillo-corteconti/grillo-corteconti/grillo-corteconti.html}, accessed 2 July 2009.}. In the following three weeks, up to 7 August, no news sources discussed V-day, while blogging activity remained constant, at an average of four to five blog posts per day. On 8 August, the official anthem of V-day, authored by Italian songwriter Leo Pari, was released. The song was advertised on Grillo's blog and ``re-blogged" in the following days, by many bloggers supporting V-day. Italian music magazine \emph{Rockstar} also reported the release of the V-day anthem.

Besides the release of the V-day anthem, no other major theme can be deduced from our blog and news content analysis for the month of August. In fact, from 8 to 30 August, news items were very sparse and were mostly V-day endorsements from small, local, online newspapers, following a post from Grillo (``V-day: an instruction manual") that pleaded for promotion and endorsement of the rally via the Web\footnote{V-day's instruction manual, at \url{http://www.beppegrillo.it/2007/08/vday\_istruzioni.html}, accessed on July 1 2009.}. The blogosphere activity maintained a constant rhythm of about six blog posts per day, with a clear prevalence of blog posts in favor. Many blog posts published in this period dealt with logistical aspects of the organization of V-day, \emph{e.g.}, some blogs reported on the activity of MeetUp.com groups and procedures to secure municipal authorization for the rally. Towards the end of the month, however, blogging events began to rise. On 29 August, politician and former judicial prosecutor Antonio Di Pietro formally endorsed the V-day rally on his blog\footnote{Antonio Di Pietro endorses V-day, at \url{http://www.antoniodipietro.com/2007/08/otto\_settembre\_parlamento\_puli.html}, accessed 15 June 2009.}. Two days later, \emph{La Repubblica} ran a story entirely dedicated to V-day\footnote{First news article entirely dedicated to V-day, on \emph{La Repubblica}, at \url{http://www.repubblica.it/2007/08/sezioni/cronaca/grillo-v-day/grillo-v-day/grillo-v-day.html}, accessed 16 May 2009.}. Although the article did not appear on the front page of the printed newspaper, it briefly appeared on the main page of the online version on 1 September\footnote{September 1 snapshot of Repubblica.it homepage cached on the Internet Archive, at \url{http://web.archive.org/web/20070901062315/http://www.repubblica.it/}, accessed 16 May 2009.}. This was the only relatively visible news story about V-day published by a mainstream newspaper prior to the rally.

The endorsement by Di Pietro and the article by \emph{La Repubblica}, just days prior to V-day rally, stirred both the social and mainstream media: the average volume of daily blog posts increased by a factor of five in a week, reaching over 50 posts in one day on 7 September. During this period, blog posts were overwhelmingly in favor of V-day. In particular, many of these posts focused on logistics, \emph{e.g.}, announcing exact dates and places of the protest in various cities, and publishing details and reports of dedicated MeetUp.com events. During this week, however, a more substantial number of blog posts opposed to V-day were also published. It is interesting to note that the posts during this period mostly fall into two categories: those that critique the controversial figure of Grillo, who is portrayed as a demagogue, and a populist leader\footnote{For example \url{http://www.cronopio.info/?p=401}, accessed June 30, 2009}, and those that direct their criticism specifically to the law reforms at the core of the V-day protest\footnote{For example \url{http://www.terzoocchio.org/controinformazione/il-testo-della-legge-proposta-da-beppe-grillo/2007/09/}, accessed 30 June 2009.}. During this week, V-day also received, for the first time, wider coverage in news sources, surpassing an average of five news items per day. However, for the most part, all the stories published during the period prior to V-day came from local, small-scale venues with very low readership (news volume in favor of V-day reached a high of 12 entries on 7 September). In other words, besides the dedicated story published (online) by \emph{La Repubblica} on 31 August, mainstream news sources were virtually silent until V-day occurred.

On 8 September V-day took place, with crowds filling the streets of Bologna and over 200 other cities in Italy and 30 abroad. The number of blog posts published on that day rose to 72. Nearly all of them were in favor of V-day. In particular, from our content analysis we note that a large portion of posts published on that day were posted after the demonstration took place: bloggers talked about their impressions and their experiences alongside photos, videos and audio clips taken throughout the day\footnote{According to a blog post updated on 11 September 2007, over 3,000 photos and 2,000 videos tagged with ``vaffaday" were uploaded to Flickr and YouTube, respectively, following the demonstration. Source: \url{http://vday.wordpress.com/2007/09/11/i-numeri-del-v-day/}, accessed 30 June 2009.}. Very importantly, many bloggers criticized the total lack of coverage of the event by the mainstream media. For example, blogger Daniele Martinelli called major TV broadcasting channel on 7 September asking them whether they intended to run a story on V-day only to hear that they were either not aware of the protest or they said that they would not cover it\footnote{A video of Daniele Martinelli calling all major TV news channels is available on YouTube, at \url{http://www.youtube.com/watch?v=BiidA9a3qtg}, accessed July 1 2009}. As a matter of fact, with the exception of RAI 3, V-day was not mentioned at all in the mid-day and afternoon news editions of major national TV channels, \emph{i.e.}, while the demonstration was taking place\footnote{Some bloggers watched all major news editions broadcasted on 8 September and documented coverage of V-day, or the lack thereof. An example is the comment by Giuliano Artefelli (9/9/2007) on Alessandro Ronchi's blog, at \url{http://www.alessandroronchi.net/2007/v-day-il-giorno-dopo/}, accessed July 9 2009.}. By the early evening, however, more ample coverage was given to V-day by some online mainstream newspapers such as \emph{Il Corriere, La Stampa} and \emph{La Repubblica}. Only \emph{La Repubblica}, however, led with an extensive story on V-day (both in the paper and online versions)\footnote{Although we could access exact date and time of publication of all news items, we could not always obtain the homepage of the online papers on which the stories were published, so that we could not determine how visible these stories were. For some sites, we used the Internet Archive's Wayback Machine to access cached versions of the papers' homepage. However, only \emph{La Repubblica} gets cached daily. Other popular newspapers, such as \emph{Il Corriere della Sera} and \emph{La Stampa} are only cached once or twice a week.}.

Although V-day eventually earned some visibility in mainstream newspapers \emph{a posteriori}, it is interesting to note that there was a crucial discrepancy between the themes discussed in the blogosphere and by the mainstream news stories in the hours following the conclusion of the event. Bloggers were mostly concerned with a) uploading content taken at the demonstration (audio, video and photos); b) sharing their views on the successful outcome of the demonstration; c) discussing the huge turnout of young people at the rally; and, d) discussing the role and importance of the Internet and social media for the organization of new forms of political participation. Most newspapers, on the other hand, ran very short stories focusing mostly on a) the derogatory name of the demonstration; b) Grillo's inflammatory attacks to corrupted politicians in his show in Piazza Maggiore; and, c) the ``war of figures", \emph{i.e.}, disputes over estimating the exact number of participants in the protest. The major motive of the protest --- the request for a new law banning criminal convicts from Parliament --- was largely ignored by mainstream media, or only mentioned \emph{en passant}.

Only two days after V-day (on 10 September) more news articles appeared --- the overall volume of news items reached its highest peak on this day, with 13 stories neutral to V-day and five opposed, nearly all published by mainstream newspapers. These stories were more in-depth analyses and commentaries. Some of them were authored by press editors and well known, experienced journalists. Again, there was very little focus on both the reasons behind the political rally and the novel form of information generation and dissemination by which the protest was organized. Rather, the themes that emerged from these articles revolve around the figure of Grillo, portrayed as a social buffoon, a populist, a demagogue, and an initiator of anti-political movements\footnote{We compiled a list of some of the adjectives attributed to Grillo by the press in the days following V-day. For reasons of space we cannot publish the entire list with the sources. We limit ourselves to some notable examples (in Italian): \emph{polemico, ridicolo, forcaiolo, mistificatore, populista, qualunquista, nuovo Mussolini, illiberale, ignorante, antipolitico, irriverente, buffone, inutile, provocatore, imbarazzante, esagerato, incoerente, guru}}. These series of critiques stirred the ``opposed" side of the blogosphere that, for the first time, reached a volume of 10 posts per day for three days in a row (10, 11 and 12 September). The themes discussed by bloggers against V-day reflected very well those found in the mainstream newspapers, suggesting that there was a considerable interplay between the blogosphere and the media --- the themes of ``populism" and ``demagogy" first appeared in some blogs (about a week before V-day); later, were fomented and aggrandized by mainstream media; finally, they became the core of discussion for the community of bloggers opposed to V-day (and more broadly, to Grillo).

In sum, from the above presentation the following picture emerges. V-day was organized, supported and disseminated solely by the initiative and effort of bloggers and small, local news agencies. During its organization, V-day received very little critique and virtually no coverage from large mainstream media sources. The first news articles about V-day with substantial visibility appeared in the online press only the day after the rally was concluded; extensive commentary and critique followed. Moreover, we note a thematic interplay between mainstream media and the blogosphere. Although we did not study this correlation statistically, we find from our content analysis that themes that were covered by mainstream media were then subject to intense discussion and critique by bloggers, while more underground themes that were born in the blogosphere were generally ignored by the media. Moreover, as discussed above, the press shifted the discussion away from the core motives of the demonstration, by focusing its criticism around the controversial figure of Grillo. Grillo was portayed as an agitator who uses social media and other Web-based participatory platforms to persuade protesters (labeled by the media with a rather diminutive ``grillini", or ``little Grillos"). The criticism by the mainstream media, therefore, mostly focused on the figure of Grillo and the environment in which the political protest emerged. In the next section, we attempt to deepen our understanding of this environment and its relationship to the central figure of Grillo, by analyzing the structure of the V-day blog network.

\section{Part 2: Link analysis of the V-day blog network}

\subsection{Data and methods}

From the previous section, a dissonant thematic relationship between blogs and the mainstream media emerges. In particular, we find that the overall critique of many news outlets after V-day was directed at the methods of information production and dissemination employed by Grillo. The papers assert that Grillo prides himself on using the open, bottom-up platforms of social media, but given his peculiar status in popular culture (and thus, the blogosphere), his information dissemination strategy resembles the typical broadcast nature of television and other traditional media.

With this part of the study we explore the role of Grillo in the organization of V-day from a network analysis perspective. We have found that the foundation upon which V-day was organized, supported and disseminated is the Italian blogosphere. The blogosphere encompasses, by definition, not only the ensemble of blogs, but also the interconnection among them. In other words, the power of blogging and other emergent forms of Internet-based participation on social media lies in their network nature, \emph{i.e.}, in their interlinking structure. We were curious to explore the network structure of the blog environment involved in the organization of V-day to understand its typology and the role of Grillo, from a network analysis standpoint.

Although various layers of interlinking among blogs exist, a typical practice for many bloggers is to keep a blogroll, \emph{i.e.}, a list of frequently read blogs and sites. Looking at this layer of interlinking, it is possible to construct a directed network in which nodes represent blogs (URLs) and directed edges represent hyperlinks (\emph{e.g., blog A links to blog B} is represented by an arrow from URL A to URL B). We constructed such a network using the blog post corpus discussed in the previous section, with the intent to measure the nature of interconnection in the community of bloggers that discussed V-day.

The initial population set was the blog corpus discussed in the previous section, consisting of all valid, manually filtered URLs pointing to posts about V-day. We converted this set to a list of URLs pointing to homepages (rather than single posts), and later eliminated all duplicates and inconsistencies. After adding Grillo's blog (that was excluded from the first part of the study), we ended up with a population size of 315 URLs (the nodes of the network). We then created a script to automatically a) visit every (source) URL in our list; b) harvest the blogroll\footnote{By and large, blogrolls are found in blog homepages (very often in the sidebar, along with blog archives, recent posts, etc.). In other cases, blogrolls can be found on a dedicated page. Some bloggers do not maintain a blogroll at all. We performed manual and automated checks to make sure that the collected data was as accurate and complete as possible}; c) inspect every (target) URL in the blogroll; and, d) create an edge from the source to the target URL if the target URL is present in our population set (\emph{i.e.}, if the target blog has discussed V-day).

Before we step into the presentation of the network analysis results, it is important to state both an assumption and a limitation associated with this part of the study. The assumption we make here is that blogroll linking reflects readership and information consumption. We do not and cannot know what information actually gets read and consumed by bloggers. But, for the purpose of the current presentation, we assume that if blogger A lists blogger B in their blogroll, then blogger B is probably an information source for blogger A. In other words, blogger A is exposed to information produced by B. Thus, if a blogger links to Grillo's blog, we can assume that a) they are aware of Grillo; b) they follow Grillo on a regular basis; and, c) that they would hear Grillo's announcement of an event, such as V-day. Based on this assumption, network centrality (\emph{i.e.}, the number of incoming links of a node) becomes a form of information influence.

This part of the study is also subject to a limitation. The network that we present below was constructed using the blogroll interlinking structure of the V-day blogosphere, as computed on 13 May 2009 --- roughly 1.5 years after V-day took place. This means that the network we study today is likely to be very different from the network of interlinking and readership on 8 September 2007. It would have been appropriate to perform this kind of analysis longitudinally as well --- to detect specific alterations in network structure with time, throughout the organization of V-day and after. Studying the network at its present state, however, enables us to draw conclusions \emph{post V-day}, based on a fairly mature, and possibly stable, state of the Italian blogosphere.

\subsection{Findings}

Using the procedure presented above, we constructed an interlink network of the blog environment involved in discussions about of V-day (including all blogs: in favor, neutral, and opposed). Table 5 contains the 16 top blogs in this network, sorted by in-degree centrality. The in-degree centrality is the number of incoming links that a blog received, or the number of times that it was cited in the blogrolls analyzed. Besides in-degree centrality, the table lists labels (which are employed in the following figures), the blog's overall opinion towards V-day (calculated as explained in Part 1 of this article), and the overall Technorati Web ranking, as of 14 May 2009\footnote{Some notes about Table 5. Piero Ricca (\url{http://www.pieroricca.org}) has been critical of Beppe Grillo in many occasions, but participated and supported V-day (\url{http://www.pieroricca.org/2007/09/09/ieri/}) and later events organized by Beppe Grillo (\url{http://www.pieroricca.org/2007/09/08/torino/}), and thus was classified as ``in favor". Daniele Luttazzi (\url{http://danieleluttazzi.it}) was classified as ``neutral" for he partly supported the motives behind V-day but criticized Grillo's methods in a blog post that was published a few days before V-day, but that is no longer available. Voglioscendere (\url{http://voglioscendere.it}), a blog authored by journalists Pino Corrias, Peter Gomez and Marco Travaglio, was opened on 20 September 2007, just over a week after V-day. Though the blog could not officially endorse V-day, it played a major role in promoting subsequent events launched by Beppe Grillo and for this reason was included in this study}. Looking at the Technorati ranks of blogs in Table 5, one immediately realizes that this table differs greatly from Table 3 (which lists blogs that most discussed V-day, regardless of their rank or centrality). While Table 3 was largely populated with sole-authored, low visibility blogs, Table 5 contains many popular blogs (nearly half of the sites listed are in Technorati's Top 10,000).

\begin{table*}[htp]
\centering
\begin{footnotesize}
\begin{tabular*}{\textwidth}{@{\extracolsep{\fill}}lccrr}
\textbf{URL}&\textbf{label}&\textbf{in-degree}&\textbf{opinion}&\textbf{rank}\\
\hline\hline
http://www.beppegrillo.it/&A&133&in favor&58\\
http://www.danieleluttazzi.it/&C&31&neutral&9,894\\
http://www.voglioscendere.it/&D&31&in favor&1,627\\
http://www.pieroricca.org/&B&24&in favor&9,458\\
http://www.uaar.it/&G&17&in favor&10,007\\
http://www.webgol.it/&F&14&neutral&27,209\\
http://alessios4.blogspot.com/&E&11&in favor&17,725\\
http://guerrillaradio.iobloggo.com/&H&11&in favor&5,821\\
http://www.downloadblog.it/&I&10&in favor&9,363\\
http://www.ecoblog.it/&J&10&in favor&10,218\\
http://www.francarame.it/&K&8&in favor&72,600\\
http://www.jacopofo.com/&L&7&in favor&15,156\\
http://odiostudioaperto.blogspot.com/&M&7&in favor&53,734\\
http://www.canisciolti.info/&N&6&neutral&-\\
http://www.ilpassatore.it/&O&6&in favor&209,741\\
http://elblogditeo.blogspot.com/&P&6&in favor&46,819\\
\end{tabular*}
\end{footnotesize}
\caption{Top sites in the V-day blog network, sorted by in-degree centrality}
\end{table*}

In Table 5, Beppe Grillo's blog is by far the most linked blog in the network, with an in-degree of 133, \emph{i.e.}, about a third of the blogger population links to Beppe Grillo. The high in-degree centrality of Grillo's blog reflects the popularity and influence that Grillo exerts on the blog network under study. This results, however, is not particularly surprising for two reasons: first, because we are focusing here on a blog network built around V-day --- an event launched by Grillo himself; second, because Grillo has been the most popular and linked Italian blogger for years. In fact, even though two-thirds of the bloggers did not directly link to him, they probably are active followers of his blog in any case --- given his popularity.

The more interesting result that emerges from Table 5 is the fact that a number of other blogs receive a fair amount of blogroll citations. Examples are the blogs of comedian Daniele Luttazzi and political blogger Piero Ricca as well as the journalistic blog Voglio Scendere. This suggests that Grillo is not the sole hub for information transfer and dissemination within the V-day blog network, \emph{i.e.}, the network is not a typical broadcast network. Other (smaller) information hubs are present, making the network heterogeneous in its composition. In Figure 4, we visualize the entire V-day blog network to highlight more in detail its topological features.

\begin{figure}[h!]
	\centering
		\includegraphics[width=1.0\textwidth]{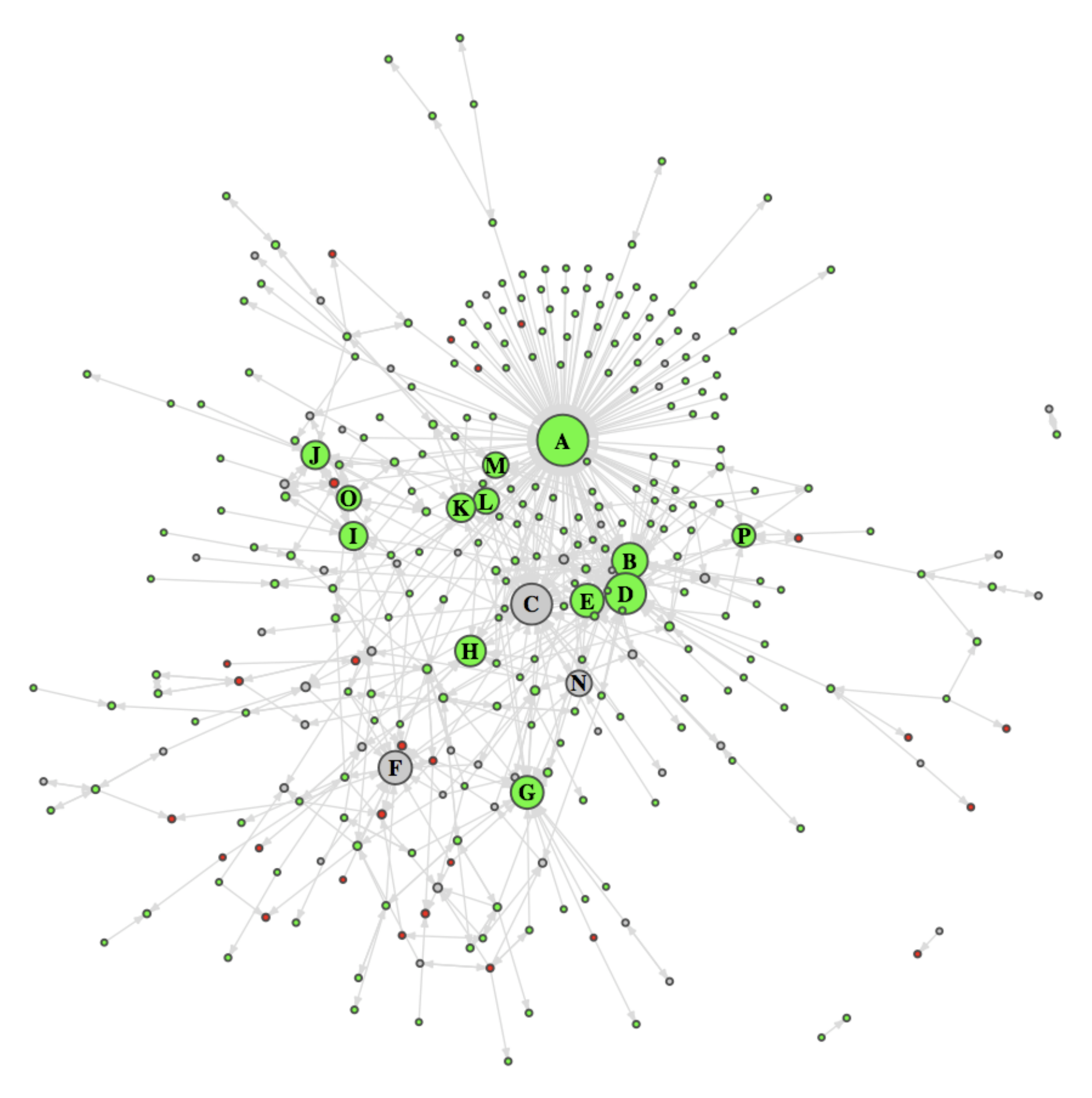}
	\caption{Interlink network based on the blogrolls of the blogs discussing V-day, including Grillo's blog (node labeled ``A")}
	\label{fig:4}
\end{figure}

\begin{figure}[h!]
	\centering
		\includegraphics[width=1.0\textwidth]{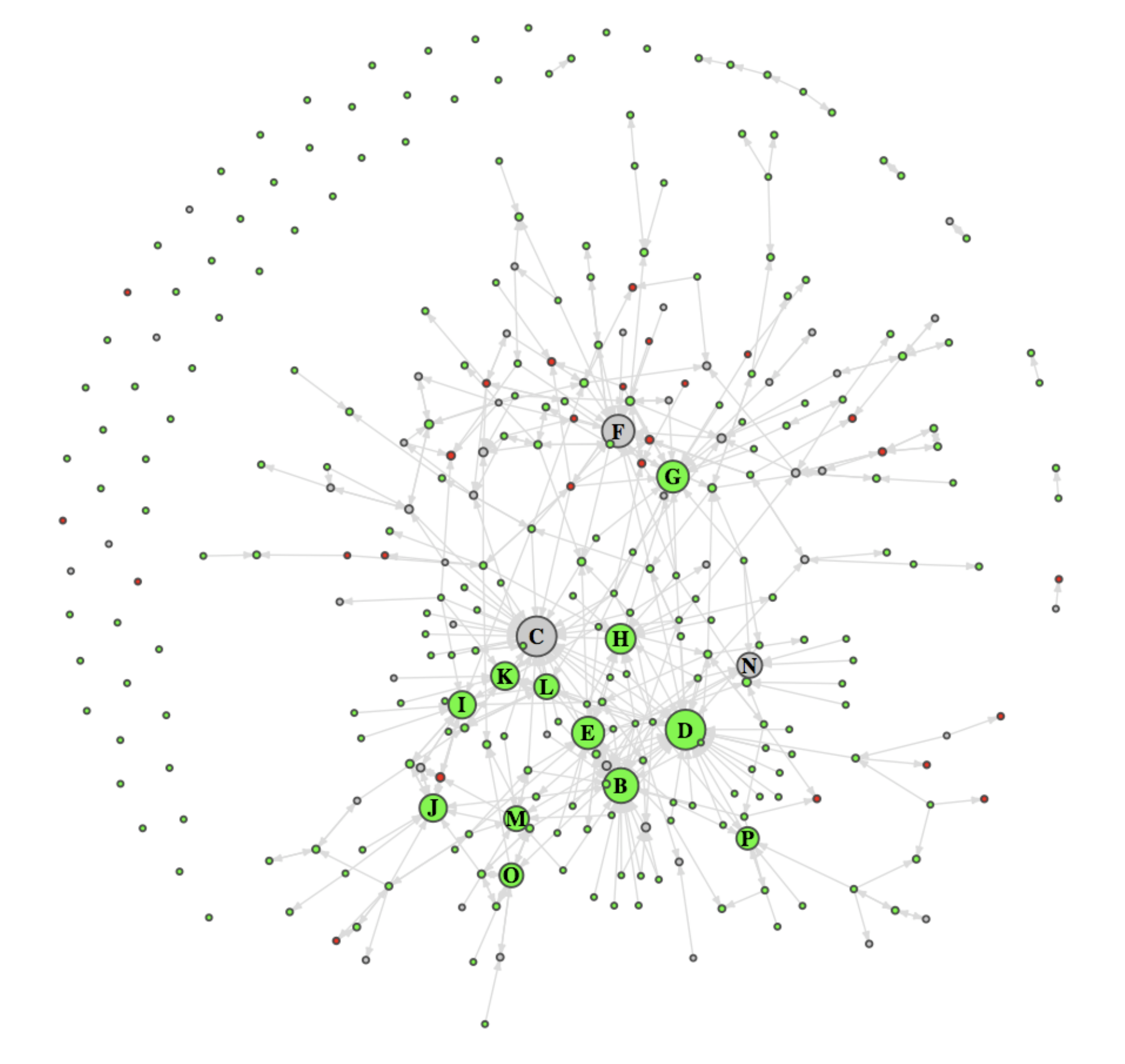}
	\caption{Interlink network based on the blogrolls of the blogs discussing V-day, excluding Grillo's blog}
	\label{fig:5}
\end{figure}

In Figure 4, nodes represent blogs that discussed V-day and directed edges represent hyperlinks between them. Node diameter is proportional to the in-degree centrality of each blog. Node colour reflects the position of the blog towards V-day event (green: in favor, gray: neutral, red: opposed). The 16 top blogs of Table 5 have been labeled with the letters indicated in Table 5. Eyeballing the network of Figure 4, one realizes that the position of Grillo's blog is indeed very prominent. This is not only because it has a very large in-degree centrality, but especially because just about half of its blogroll citations (precisely 68, depicted in the upper part of the diagram) come from nodes that are disconnected from the rest of the graph. Building on the assumption presented above, we can thus speculate that these isolated blogs only heard about V-day through Grillo's blog, in a broadcast fashion. They are not part of the broader interlinking structure that characterizes the rest of the V-day blogosphere. Despite an abundance of small nodes (lowly cited blogs), the network is fairly heterogenous, with many different node centrality levels. This image contradicts the hypothesis that Grillo functions as a top-down broadcast-like channel of information transfer. We were curious to push this finding further and isolate the influence of Grillo's blog on the function of the network. We assume here that the primary function of the network is to disseminate information about V-day and we ignore the fact that V-day was indeed first announced on Grillo's blog. In Figure 5, we present the same network of Figure 4, but with its most central node --- Beppe Grillo's blog --- removed.

As expected, part of the network is dismantled by the removal of Grillo's blog and becomes disconnected from the rest of the graph. A solid portion of the graph, however, stays connected in a giant component (the central connected component), indicating that the network is resilient to node removal. Figure 4 and 5 are best analyzed in conjunction with the results presented in Table 6. This table shows some fundamental parameters and network statistics relative to the full graph, labeled \emph{G} (\emph{i.e.}, Figure 4) and the partial network (without Grillo's blog), labeled \emph{g} (\emph{i.e.}, Figure 5).

\begin{table*}[htp]
\centering
\begin{footnotesize}
\begin{tabular*}{\textwidth}{@{\extracolsep{\fill}}lrr}
\textbf{quantity/network}&\textbf{$G$}&\textbf{$g$}\\
\hline\hline
Nodes&315&314\\
Edges&599&466\\
Connected components&4&64\\
Size of giant component&309&241\\
Average path length, $\ell$&2.746&2.870\\
Clustering coefficient, $C$&0.184&0.202\\
\end{tabular*}
\end{footnotesize}
\caption{Statistics for graphs \emph{G} (entire V-day interlinking network) and \emph{g} (network with Grillo's blog removed)}
\end{table*}

The first two rows of Table 6 display the number of nodes --- and edges in networks \emph{G} and \emph{g}. By removing Grillo's blog (number of nodes goes to 314), the 133 edges pointing to Grillo disappear, causing the number of edges to drop from 599 to 466. As a result, the number of connected components jumps from 4 to 64, with the majority of newly created components being lone nodes (as displayed in the outer edges of Figure 5). Despite the network being highly fragmented by the removal of Grillo, the size of the giant component (\emph{i.e.}, the largest connected subgraph in the network) is not heavily affected (it drops from 309 to 241). The presence of a giant component indicates that a solid core of loosely interconnected blogs exists, \emph{i.e.}, the vast majority of bloggers who blogged about V-day either know each other or are accessible within a short path length.

The following two measures --- average path length and clustering coefficient --- refer specifically to the topology of the network and are calculated on the giant component only. The average path length is the average shortest number of steps needed to connect any two nodes in a network and is a reliable indicator of the efficiency of information transfer in a social network (Wasserman and Faust, 1994). Since the network presented here is a blog interlinking network, the path length can be conceptualized as the number of ``mouse clicks" needed for a blogger to reach another blog. Thus, an average path length of 2.746 means that a blogger in the V-day network would learn about every other blogger who discussed V-day by clicking on less than three blogroll links (starting from from their own blogroll). The results in Table 6 demonstrate that this value is not highly affected when Grillo is removed from the network (path length only increases slightly, to 2.870).

The last value reported in Table 6, the clustering coefficient, measures the density of clique-like triangles in a network. Short average path length and high clustering coefficient are typical characteristics of many real and small-world networks (Watts and Strogatz, 1998; Newman, 2003). Our results demonstrate that the removal of Grillo has the effect of increasing, rather than reducing, the level of clustering of the network (from 0.184 to 0.202). This is probably due to the fact that Grillo was connected to a number of marginal nodes that reduced the overall clustering coefficient, rather than increasing it.

Overall, the results of this limited network analysis indicate that the V-day blog network is a small-world network that features a solid core of interconnected bloggers and is fairly resilient to node removal. In other words, this network study confirms the central role of Grillo's blog in the V-day blog network, but contradicts the hypothesis that Grillo was the only broadcast-like source of information. Rather, we find that information was transferred across heterogeneous nodes in a moderately robust core of interconnected blogs.

\section{Conclusions}

This article set out to explore the organization, promotion and public perception of V-day, a political rally that took place on 8 September 2007 to protest against corruption in the Italian Parliament. Launched by blogger Beppe Grillo, and promoted via a word-of-mouth mobilization on the Italian blogosphere, V-day brought close to one million Italians to the streets, while being mostly ignored by mainstream media. Such a lack of coverage by the mass media made V-day a special case study to investigate the role of the Internet and social media in bringing about political participation in Italy.

Our study of the V-day demonstration is in two parts. First, we analyzed the volume and content of online articles published by both bloggers and news sources from 14 June to 15 September 2007. We found that the V-day event received widespread discussion, audience and dissemination in the blogosphere, but very little coverage by mainstream newspapers, suggesting that the success and popularity of the event can be attributed to the blogosphere only. This finding is particularly relevant when one considers that public discourse is usually set by the mainstream media. Our study, however, shows that the success of V-day was made possible by the blogosphere, therefore with the public and not the mainstream media setting the agenda for collective social action. Overall, we noted that bloggers were mainly in support of the event throughout its organization, while news media were largely neutral and covered the event only after it took place. Also, we noted that the majority of bloggers and news sources in favor of the event were, respectively, regular citizens and small-scale local outlets, suggesting a strong grassroots component in the organization of the demonstration. While our analysis suggests that V-day was mobilized primarily on the blogosphere, we found a unidirectional interplay between the mainstream media and the blogs: themes covered by the mainstream media were intensely discussed by bloggers, although, conversely, mainstream media rarely followed bloggers' discussions, showing the insulated, closed nature of the mainstream media ecology in Italy. This finding suggests that although the blogosphere was pivotal in promoting the political mobilization of citizens for V-day, it did not succeed in influencing the agenda of mainstream media. This lack of success of social media in influencing mainstream media is mostly significant when one considers that the Internet is increasingly becoming an environment for discussion and sharing of political ideas, while mainstream media are undergoing an intense period of turmoil, with strong pressures towards censorship of current political events.

In the second part of the article, we study the interlinking structure of the blogosphere network in the context of V-day, to determine its structure, and in particular its levels of heterogeneity and resilience. Our blogroll link analysis of the V-day network contradicts the hypothesis that Grillo functioned as a top-down broadcast-like channel of information transfer for the organization and promotion of V-day. Despite the prominent position of Grillo's blog in the network under study, we find that the V-day blogosphere is heterogeneous, \emph{i.e.}, with many different nodes of all levels functioning as information transfer hubs, and resilient, \emph{i.e.}, the removal of its most central node (Grillo's blog) does not dismantle it.

We understand that the analysis presented in this article only scratches the surface of the nature of the relationship between bloggers and mainstream media in bringing about political mobilization. Indeed, with this article, we do not attempt to portray in full the Italian blogosphere and its interplay with the national media ecology: we focus on the bloggers' linking behavior in the context of a specific political event. Future research should look in further detail at the broader interplay and linking structure that exists between the Italian blogosphere, as a whole, and mainstream media sources. Yet, we argue that our case study of V-day is sufficient to uncover a marked dissonance between the Italian mainstream media and novel forms of political participation emerging on the Web, which, we speculate, is the result of collusion between media and politics and limited freedom of press. It is this alarming media and political landscape that we hope to bring to the attention of Italian and international scholars.

In conclusion, we would like to reiterate the most important finding of this article: V-day was Italy's first experiment of political organization and action developed primarily via the use of social media on the Web. The popular mobilization seen at the V-day rally not only points to the effectiveness of the Internet as a platform for political organization. It also highlights the importance of taking into consideration the context and the media ecology in which these novel forms of grassroots organization take place. In the case of Italy, a country with a consolidated television culture and one of the lowest Internet penetration rates in the European Union (50.1 percent; the average in the EU is 63.1 percent)\footnote{Internet penetration data obtained from \url{http://www.internetworldstats.com/}, accessed 29 August 2009}, the mobilization of hundreds of thousands regular citizens via online social media has no precedent. In this context, we wonder whether the organization of V-day, that brought nearly two percent of the Italian population to the streets, could have been possible in another country. Leaving cultural, social and political differences aside for a moment, we would like to conclude with this analogical question\footnote{In the final question, we choose Arianna Huffington only because the Huffington Post is the most followed political blog in the United States (and the world)}: \emph{could Arianna Huffington mobilize six million Americans without any support from the mainstream media?}
\newpage

\section{About the authors}

\textbf{Alberto Pepe} is a Ph.D. candidate in Information Studies at the University of California, Los Angeles and a researcher in the Statistics and Data practices group of the Center for Embedded Networked Sensing. He is an information scientist interested in the study of socio-technical systems: networks of people, artifacts, data and ideas. Prior to starting his Ph.D., Alberto worked in the Information Technology Department of CERN, in Geneva, Switzerland, where he developed digital library software and promoted open access among particle physicists. He also worked in the Scientific Visualization Department of CINECA, the Italian Scientific Consortium, based at the University of Bologna. Alberto holds a M.Sc. in Computer Science and a B.Sc. in Astrophysics, both from University College London, U.K. Alberto is from Manduria, a wine-making town in Puglia, Southern Italy. Web: \url{http://albertopepe.com/}\\

\noindent
\textbf{Corinna di Gennaro} is a Research Associate at the Oxford Internet Institute at the University of Oxford. She is a sociologist working on the social implications of Internet adoption and use for civic and political engagement. Corinna holds a D.Phil in Sociology from the University of Oxford and a BSc. in Sociology from the London School of Economics (LSE). She was a Fellow at the Berkman Center for Internet and Society at Harvard University, a Postdoctoral Research Fellow at the USC Annenberg Center for Communication and Survey Research Officer at the Oxford Internet Institute. Her main expertise is in political sociology, survey analysis and quantitative methods of social research. She is from Milan, Italy. Web: \url{http://corinnadigennaro.com/}

\section{Acknowledgements}

We would like to thank Christine Borgman of the University of California, Los Angeles and Peter Bearman of Columbia University for reading and providing comments on an earlier version of this manuscript. Many thanks to Chris Starr for giving us very detailed editing comments on the final draft. Also, thanks to Alice Mattoni of the European University Institute and Francesca Di Donato of the University of Pisa for suggesting ideas and literature on social movements and political mobilization. We are also grateful to Spencer Wolff of Yale University and Simo Bennani of Pepperdine University for discussing with us the socio-political implications of studies of this kind. Material from this article was presented at the ISA RC51 conference ``Modernity 2.0, Emerging Social Media Technologies and their impacts" at the University of Urbino, Italy on 2 July 2009, where it was awarded with the Walter Buckley Memorial Award for ``Excellence in Presenting Sociocybernetics."

\section{References}

Massimiliano Andretta, D. Della Porta, L. Mosca and H. Reiter, 2002. \emph{Global, Noglobal, New Global: Le proteste contro il G8 a Genova}. Rome: Laterza.

\vspace{2.5mm}\noindent
Samuel Best and Brian S. Krueger, 2005. “Analyzing the representativeness of Internet political participation,” \emph{Political Behavior}, volume 27, pp. 183-216.

\vspace{2.5mm}\noindent
Regina Barzilay, Noemie Elhadad, and Kathy McKeown, 2002. “Inferring strategies for sentence ordering in multidocument news summarization,” \emph{Journal of Artifical Intelligence Research (JAIR)}, volume 17, pp 35-55.

\vspace{2.5mm}\noindent
Mridul Chowdhury, 2008. “The role of the Internet in Burma's Saffron Revolution,” Harvard University, Berkman Center for Internet \& Society, Research Publication, number 2008-08, at \url{http://cyber.law.harvard.edu/publications/2008/Role\_of\_the\_Internet\_in\_Burmas\_Saffron\_Revolution}, accessed 14 June 2009.

\vspace{2.5mm}\noindent
Rachel K. Gibson, Wainer Lusoli, and Stephen J. Ward, 2005. “Online participation in the U.K.: Testing a `contextualized' model of Internet effects,” \emph{British Journal of Politics and International Relations}, volume 7, number 4, pp. 561-583.

\vspace{2.5mm}\noindent
Joshua Goldstein, 2007. “The role of digital networked technologies in the Ukrainian Orange Revolution,” Harvard University, Berkman Center for Internet \& Society, Research Publication, number 2007-14, at \url{http://papers.ssrn.com/sol3/papers.cfm?abstract\_id=1077686} accessed 14 June 2009.

\vspace{2.5mm}\noindent
Thomas Johnson and Barbara Kaye, 1998. “A vehicle for engagement or a haven for the disaffected? Internet use, political alienation, and voter participation,” In Thomas J. Johnson, Carol E. Hays and Scott P. Hays (editors). \emph{Engaging the public: How the government and media can reinvigorate democracy}. Lanham, Md.: Roman and Littlefield, pp. 123-135.

\vspace{2.5mm}\noindent
Mary Joyce, 2007. “The citizen journalism Web site `OhmyNews' and the 2002 South Korean Presidential election,” Harvard University, Berkman Center for Internet \& Society, Research Publication, number 2007-15, at \url{http://papers.ssrn.com/sol3/papers.cfm?abstract\_id=1077920}, accessed 14 June 2009.

\vspace{2.5mm}\noindent
Giovanni Navarria, 2009. “Beppe Grillo, the talking cricket,” In: Adrienne Russell and Nabil Echchaibi (editors). \emph{International blogging: Identity, politics, and networked publics}. New York: Peter Lang.

\vspace{2.5mm}\noindent
Giovanni Navarria, 2008. “Political anomalies and Web-based civil antibodies in Silvio Berlusconi's Bel Paese,” \emph{Recerca, Revista de Pensamenti I Analisi}, volume 8, pp. 173-192.

\vspace{2.5mm}\noindent
Mark E.J. Newman, 2003. “The structure and function of complex networks,” \emph{SIAM Review}, volume 45, number 2, pp. 167-256.

\vspace{2.5mm}\noindent
Stephen D. Reese, Lou Rutigliano, Kideuk Hyun and Jaekwan Jeong, 2007. “Mapping the blogosphere: Professional and citizen-based media in the global news arena,” \emph{Journalism}, volume 8, number 3, pp. 235-261.

\vspace{2.5mm}\noindent
Howard Rheingold, 2002. \emph{Smart mobs: The next social revolution}. Cambridge, Mass.: Perseus.

\vspace{2.5mm}\noindent
Clay Shirky, 2008. \emph{Here comes everybody: The power of organizing without organizations}. New York: Penguin Press.

\vspace{2.5mm}\noindent
Mike Thelwall and Laura Hasler, 2007. “Blog search engines,” \emph{Online Information Review}, volume 31, number 4, pp. 467-479.

\vspace{2.5mm}\noindent
Caroline J. Tolbert and Ramona S. McNeal, 2003. “Unraveling the effects of the Internet on political participation?” \emph{Political Research Quarterly}, volume 56, number 2, pp. 175-185.

\vspace{2.5mm}\noindent
Nikki Usher, 2008. “Reviewing Fauxtography: A blog-driven challenge to mass media power without the promises of networked publicity,” \emph{First Monday}, volume 13, number 12, at \url{http://firstmonday.org/htbin/cgiwrap/bin/ojs/index.php/fm/article/view/2158/2055}, accessed 7 December 2009.

\vspace{2.5mm}\noindent
Kevin Wallsten, 2005. “Blogs and the bloggers who blog them: Is the political blogosphere an echo chamber?” paper presented at the annual meeting of the American Political Science Association, Washington, D.C. (1 September).

\vspace{2.5mm}\noindent
Stanley Wasserman and Katherine Faust, 1994. \emph{Social network analysis: Methods and applications}. Cambridge: Cambridge University Press.

\vspace{2.5mm}\noindent
Duncan J. Watts and Steven H. Strogatz, 1998. “Collective dynamics of `small-world' networks,” \emph{Nature}, volume 393, number 6684, pp. 440-442.
\end{document}